# Metallic or insulating magnetic phases of transition-metal oxides in light of the Slater vs Mott controversy

Xiuwen Zhang[1], Jia-Xin Xiong[1,†], and Alex Zunger[1,*]

[1]RASEI, University of Colorado Boulder, Boulder, Colorado 80309, USA

## Abstract

Magnetic *d*-electron oxides usually have a long-range-ordered ground state and an above-ground-state disordered para phase. The historical Slater-Mott controversy still reverberating to date pertains to understanding the origin of the insulating vs metallic nature of these two phases. Slater suggested that the insulating character evolved from the long-range order (LRO) antiferromagnetic (AFM) cell-doubling, leading to energy band repulsion and thus to forming a band gap. The reliance on LRO to explain gapping suggests that the absence of LRO in the spin-disordered para-phase would deprive the system from band repulsion, thus predicting a metallic paramagnetic (PM) phase. This, however, conflicts with numerous current observations. The alternative Mott-Peierls-Hubbard view abandons the mean-field-like band structure view emphasizing instead the local interelectronic repulsion $U$ as a strong correlation signature of gapping both at the presence or absence of LRO. However, $U$ is not an easily measured or calculated quantity in periodic systems, a situation often leading in practice to treating $U$ as an adjustable parameter. The current work uses a parameter-free mean-field-like approach (Density Functional Theory, DFT), allowing energy-lowering symmetry breaking. This is done by considering local magnetic motifs as building blocks of the macroscopic AFM or PM phases. It is found that even when such motifs are isolated electronically from each other (i.e. vanishing LRO) by embedding them in quantum barriers (such as magnetic $NiO_6$ octahedra surrounded locally by wide-gap MgO) they exhibit an *intrinsic* band gap. Thus, the origin of the band gaps in AFM or PM phases need not be LRO or strong correlation and can be understood within the mean-field-like band structure approach with symmetry breaking.

† Corresponding author: Jia-Xin.Xiong@colorado.edu

* Corresponding author: alex.zunger@colorado.edu

## I. Introduction

*The Mott vs Slater controversy about gapping of magnetic phases of oxides:* The ground state of magnetic compounds often appears in a spin-long-range-ordered antiferromagnetic (AFM) or ferromagnetic (FM) configuration, whereas above a characteristic Néel temperature ($T_N$) it appears in a spin-disordered paramagnetic (PM) phase. A significant issue, debated for a long time [1,2], has been what causes different spin configurations including long-range ordered (LRO) ground-state AFM, FM and above-ground-state disordered PM in different compounds such as transition-metal oxides or halides to be either insulating or metallic. The importance of this question lies in the need to select materials that can be metals or even superconductors. This controversy was at a focus of a historical debate between Slater and Mott [1,2] more than 75 years ago. Slater's view [2] suggested that the existence of low-temperature *insulating* antiferromagnets such as NiO and MnO is a consequence of cell-doubling along a long-range order pattern, leading to band repulsion between the folded-in bands and thus the formation of the AFM crystal band gap. In Slater's words (1951) regarding the AFM phase [2], *"electrons move with opposite spins in different potentials… This results in a periodic perturbation of potential, with periodicity twice the atomic periodicity, and leads to a splitting of* each energy band in half, *with a gap in the middle."* As a consequence of Slater's reliance on long-range order to explain the AFM gapping, it followed that the *absence* of long-range order in the spin-disordered PM phase would deprive the system from band repulsion, thus leading to a *metallic* PM phase. However, with time it became apparent that quite a few 3d oxides that have an insulating AFM phase, also manifest above the Néel temperature $T_N$ a PM phase that is also insulating (provided that $T_N$ is not too high) such as NiO, MnO, $LaMnO_3$, $LaFeO_3$, $LaTiO_3$, $LaVO_3$, and $YNiO_3$ [3–7]. This contradiction with experiment was apparently not present in the alternative Mott-Peierls-Hubbard interpretation [8] that emphasized the *local* interelectronic repulsion *U* as a strong correlation signature of insulation both at the presence and absence of long-range order. This introduction of strong correlation as the crucial enabling factor signaled a historic turn away from the mean-field electronic band structure view of gapped magnets. However, the critical on-site Coulomb parameter *U* in extended solids is neither a routinely measured solid-state quantity, nor an easily calculated first-principles quantity. (Indeed, the value of *U* that needs to be assumed in Mott-Hubbard modelling to fit the measured effect can be dramatically different from the best *U* value known for the system at hand [9]). It seems that even three quarters of a century after this textbook debate ensued, neither the Slater nor the Mott approach can satisfactorily identify the nature of gapping of AFM and PM compounds.

*Density Functional Theory (DFT) calculations using energy-lowering symmetry breaking:* The exclusive need for strong correlation physics as a cause of an insulating response in paramagnets has been recently called into question [10,11]. It turns out that allowing energy-lowering magnetic symmetry breaking and/or positional symmetry breaking (dimerization, trimerization, disproportionation, octahedral tilting, etc.) in the mean-field-like DFT does predict both for spin-LRO ground states and for above-ground-state spin-disordered PM phases the correct metal or insulator gapping tendencies. Such symmetry-broken DFT approach treating LRO AFM configurations as well as disordered PM spin alloy using a large supercell [12] was able to correctly classify prototype classes of dual behaviors. This included the "AFM insulator and PM insulator" class of compounds (such as $LaTiO_3$ and $LaFeO_3$) [13], the class of "AFM insulator and PM metal" (such as $NaOsO_3$) [14], the class of "AFM not detected and PM persistent metal" (such as $SrVO_3$ and $LaNiO_3$) [13,15], and the class of "AFM not detected and PM insulator" (such as $Nb_3Cl_8$) [16]. This implies that long-range order is not crucially needed as an enabling condition to generate the insulating gap ground state, and that gapping in PM phases may not require exclusively strong correlation although it has been demonstrated in DFT-initiated dynamic mean-field theory (DFT-DMFT) that insulating paramagnets can be generalized theoretically without long-range order [17].

*What is missing:* Although the above-noted symmetry-broken DFT approach can classify real compounds into AFM and PM classes of insulators and metals, an understanding of such trends in terms of basic building block properties seems to be lacking. Specifically, one would like to determine what are the motifs present in spin-LRO and spin-disordered phases that deterministically control their metal or insulator character. In the present work we determine the structural and magnetic local motifs that carry an intrinsic band gap (with finite or zero value) so that packing such building block motifs either with LRO or disordered patterns produces the macroscopic gap.

*Construction of isolated magneto-structural motifs:* One can define for FM configurations a "single spin-motif" (e.g. in NiO a single spin-polarized $NiO_6$ octahedron, Fig. 1 upper row) and for AFM configurations a "double, opposite-spin-motif" (e.g. in NiO two opposite-spin-polarized $NiO_6$ octahedra, Fig. 1 bottom row). A *bulk* FM or PM compound is created by assembling closely packed (N=0 in Fig. 1) interacting "single spin-motifs", whereas a bulk AFM compound is created by assembling closely packed (also N=0) interacting "double, opposite-spin-motif". The respective bulk band gaps, either being metallic or insulating, are obtained by conventional (spin-polarized) DFT band structure calculation of these configurations having closely packed periodic motifs. *The key problem is to determine if in isolation such local motifs carry an intrinsic band gap, or whether the bulk band gap is a consequence of the inter-motif interaction in a given bulk geometry.*

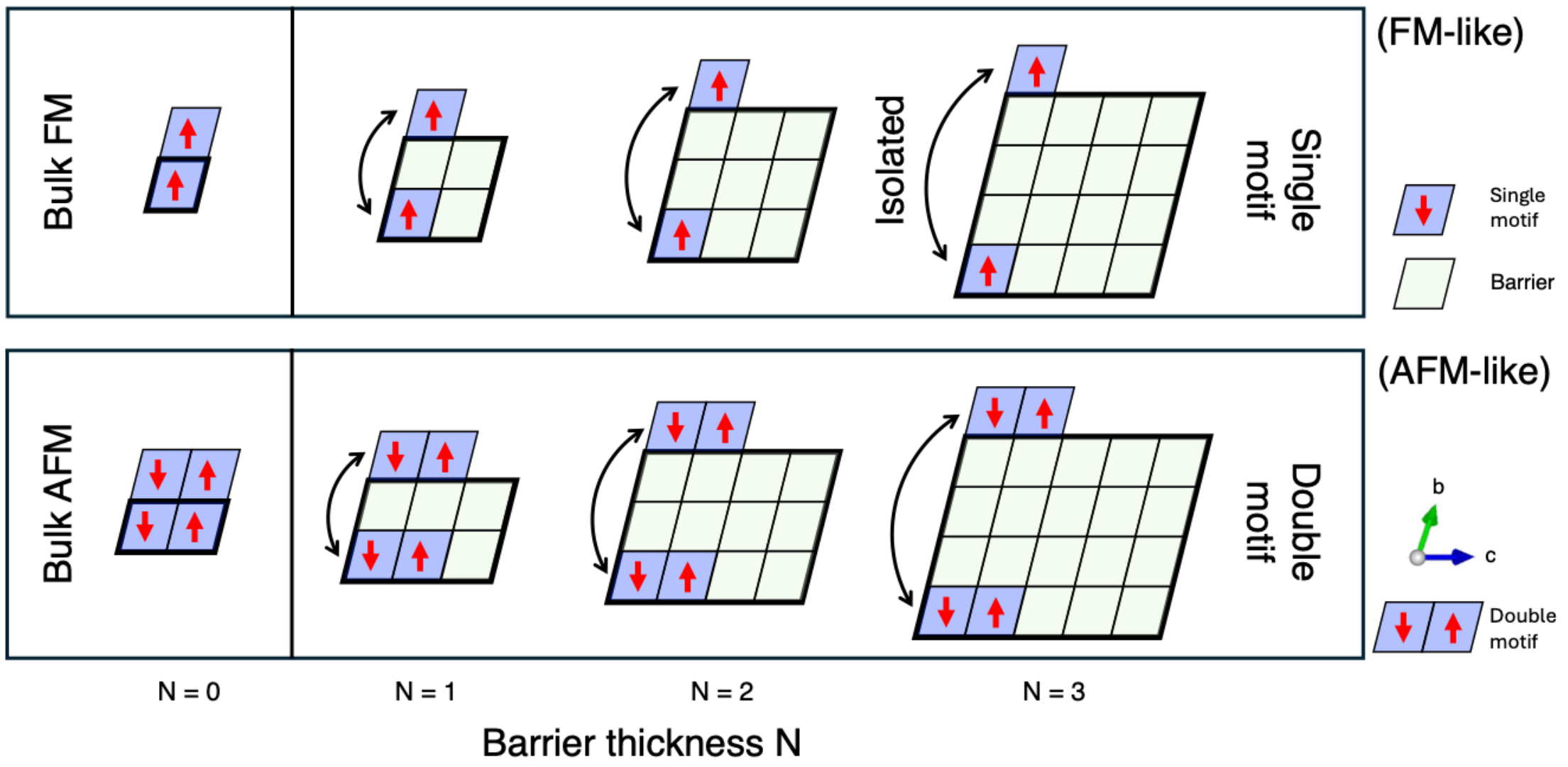

**Figure 1.** Supercell description of bulk FM (top row) and bulk AFM (bottom row) where the single spin motif [building block of ordered FM and disordered paramagnetic (PM) configurations] is made of a single spin-polarized $NiO_6$ octahedron, whereas the a "*double, opposite-spin-motif*" (building block of AFM configuration) is made of two

opposite-spin-polarized $NiO_6$ octahedra. Shown are supercells within which there is a relatively thick barrier (a Large *N*) giving rise to isolated building blocks, whereas no barrier (*N* = 0) means bulk phases (FM, AFM, or PM).

*The computational idea:* One can imagine creating "electronic vacuum" between individual motifs (larger N in Fig. 1). The computational idea is simple as following: we place between different "single spin-motif" (for FM) or between "*double, opposite-spin-motif*" (for AFM) a wide-gap nonmagnetic compound that is lattice matched to NiO of thickness of N primitive cells. We then study the electronic structure of such AFM supercells by artificially reducing the coupling between different magneto-structural motifs (using thicker barrier; larger N values in Fig. 1). We chose for the barrier material rock-salt MgO creating quantum confinement of the NiO states by the MgO wide gap barrier. To examine if quantum confinement was accomplished, we observe the eigenvalues of NiO as a function of the thickness N of the MgO barrier (Figs. 2 and 3).

To determine if *isolated* spin motifs, packed with sufficiently thick "electronic vacuum" between such individual motifs already possess a motif-intrinsic band gap or not, we borrow an idea from nanostructure quantum confinement physics [Fig. 2(a)]. Here we select a "barrier compound" (here, MgO) that has a much larger gap than NiO—the latter acts as a "quantum well" compound. Whether this model is appropriate is decided by the DFT supercell calculation where NiO octahedra are surrounded by different thickness on MgO barrier. Orbital analysis of the nature of the confined electrons in the conduction band minimum (CBM) and confined holes in the valence band maximum (VBM) is given in Fig. 2(b) and Fig. 2(c), respectively. Clearly, the CBM [Fig. 2(b)] contains Ni-d orbitals in the NiO well, whereas the VBM [Fig. 2(c)] contains O-p orbitals (from both NiO and MgO). Thus, the important NiO-d ($e_g$) orbitals that control the presence or absence of a motif band gap are confined by the MgO barrier into the NiO well.

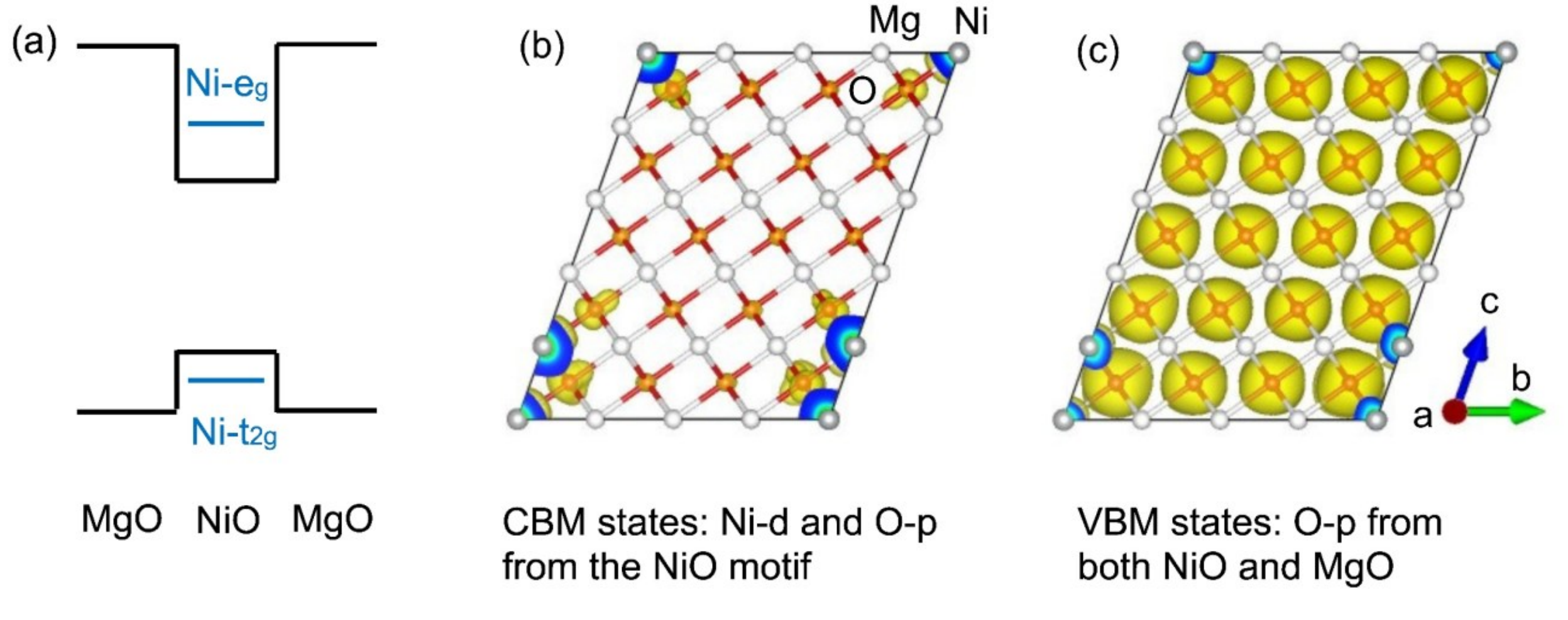

**Figure 2.** (a) Schematic of band alignment of NiO-MgO. (b) and (c) show the DFT-calculated wavefunction squared of the isolated motif (N=3 double, opposite-spin NiO motif) for (b) conduction band minimum (CBM) and (c) valence band maximum (VBM).

The *surprising* result (to be discussed with more details in Fig. 3) is that using the mean-field-like DFT, such isolated, practically non-interacting spin motifs already manifest an intrinsic band gap—independent on long-range order or disorder of such local motifs or strong correlation (which DFT lacks). Indeed, we find that using appropriately designed spin geometries as in Fig. 1, the individual isolated local motifs have an i*ntrinsic* band gap, (even larger than that of interacting

motifs–the bulk AFM phase). This suggests that the origin of band gaps in AFM, FM and PM phases is neither long-range order (Slater mechanism) nor strong correlation (Mott mechanism), but the individual building blocks of such configuration have their own intrinsic gap before being assembled into LRO or disordered configuration. These ideas including the explanation of the origin of the intrinsic gaps of the isolated motifs and results are discussed next.

## II. Method of calculations

We constructed bulk NiO and NiO-MgO structures based on the Fm-3m rock-salt lattice. The lattice constant reported for the FM phase [18] (a=4.17 Å) was used to compare different magnetic configurations at a common lattice constant. The FM-like single spin-motifs were constructed by assigning parallel magnetic moments to the Ni sites. For comparison, the AFM-like double, opposite-spin-motifs were arranged in alternating chains along the *b* axis of the rock-salt primitive cell, as illustrated in Fig. 1. For the bulk AFM phase, we used the experimentally reported magnetic structure and lattice constant [19] (a=4.17 Å), which is close to the FM phase. The PM phase of bulk NiO was represented by a spin-disordered supercell containing 160 atoms. Equal numbers of spin-up and spin-down Ni sites were distributed following the special quasi-random structures (SQS) protocols [12]. The lattice vectors and cell-internal atoms are fixed without relaxation.

We perform spin-polarized density-functional theory (DFT) calculations using the Vienna Ab Initio Simulation Package (VASP) [20,21]. We use the projector augmented-wave (PAW) method [22] and exchange-correlation functional of RPBE [23]. We use the Ni and O_s pseudopotentials for valence electrons in the PAW datasets. The plane-wave kinetic-energy cutoff was set to 368 eV. We set the converging criteria of electronic self-consistency to $10^{-6}$ eV.

## III. Analysis of the mechanisms leading to gapping in AFM, FM and PM phases

### A. Controlling the interactions between individual spin motifs

Taken together, Fig. 2 supports the needed electronic isolation of Ni-d orbitals between NiO motifs, while allowing hybridization with the surrounding O-p states. We therefore examine whether a finite, intrinsic gap persists upon motif isolation and how it relates to the local crystal-field splitting and spin polarization of the Ni-d states [24,25], irrespective of their packed as bulk LRO or disordered configuration. Fig. 3 will compare the band gaps of embedded spin motifs with those of the spin-ordered AFM and FM, and spin-disordered PM bulk configurations.

### B. Electronic properties as a function of the coupling between individual spin motifs

Fig. 3(a) shows that as the separation (barrier thickness N increasing) between two AFM building blocks (blue line) or between two FM building blocks (red lines) in a periodic supercell increase, the respective band gaps increase. This reflects the energy repulsion between CBM and VBM due to increasing quantum confinement. The saturation of the FM and AFM gap values, when the separation between the building block motifs is large enough, represents the ideal value of intrinsic band gap of *isolated* FM and AFM building blocks. Surprisingly, when the barrier thickness is large enough, both the FM and AFM band gaps of the isolated building blocks saturate to almost identical asymptotic values. This reflects the fact that the coupling between nearest opposite-spin $NiO_6$ octahedra has negligible effect, whereas the inter-coupling between two motifs separated in different supercells has an obvious impact on reducing the intrinsic band gap. Conversely, at N=0 where the building block motifs are fully coupled, one obtained the bulk band gaps of FM and AFM, where the PM gap is intermediate.

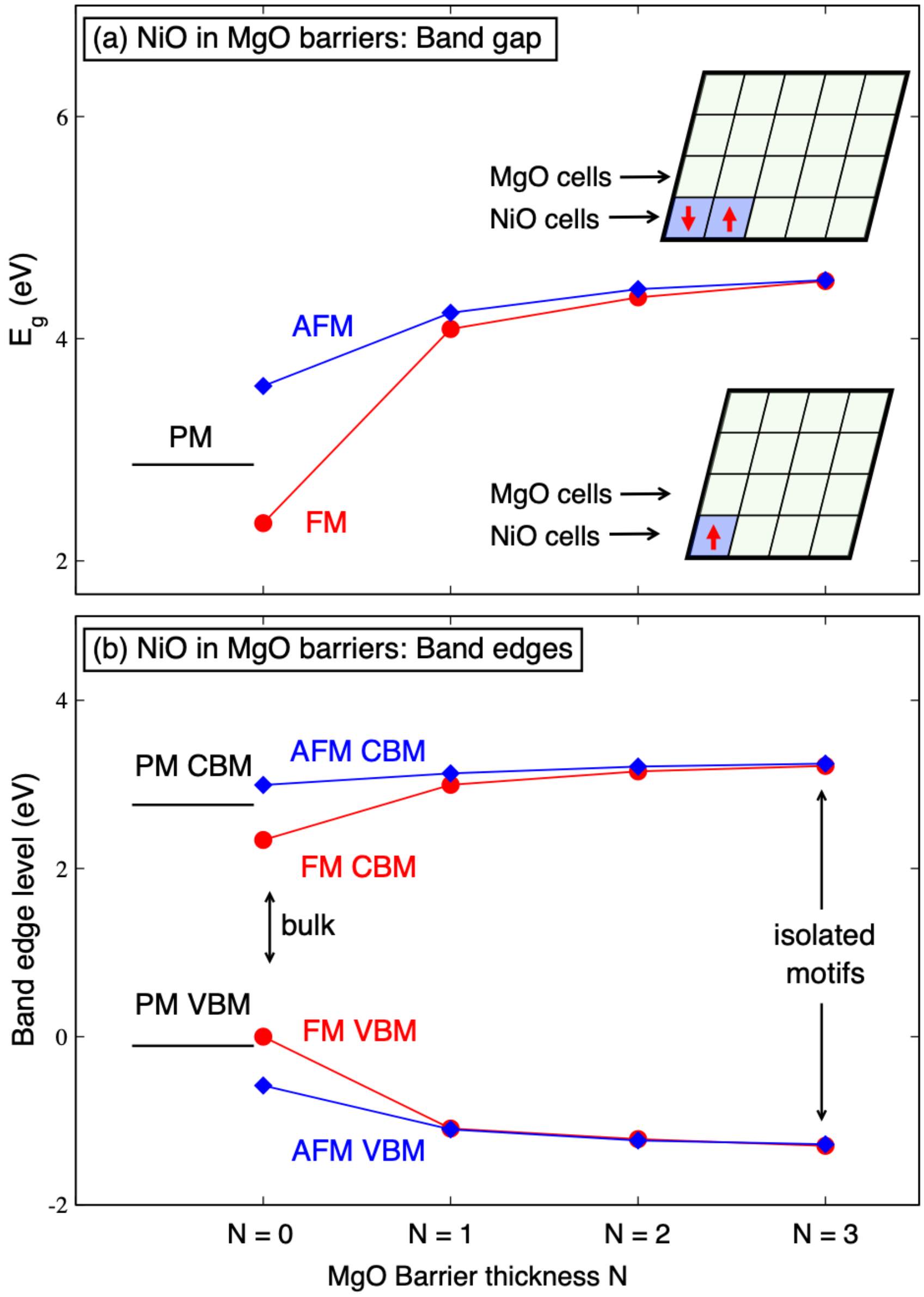

**Figure 3.** (a) Band gaps and (b) band-edge energy levels of single motif (FM-like), double motif (AFM-like), as well as spin-SQS PM configurations as a function of the thickness N of the MgO barrier layer (see Fig.1). The results are obtained from DFT spin polarized supercell calculations.

Fig. 3(b) shows the calculated band edges of *isolated* NiO building block motifs (N=1, 2, 3) and bulk AFM NiO where the building block motifs are interacting (N=0). These eigenvalues were aligned by assuming a common core O 1*s* levels. We find in Fig. 3(b) that for thick enough barrier, the band-edge energies tend to saturate and that the emerging isolated local motifs carry their signature intrinsic band gaps. The band gaps [see Fig. 3(a)] of such AFM-like spin motifs are considerably larger than the band gap of the bulk AFM structure as the inter motif coupling in the AFM structure reduces the band gap. Therefore, our calculations are opposed to the view that LRO is the reason for the band gap opening in the AFM phase.

### C. Local magnetic moments of spin motifs and bulk phases

Figure 4 shows the DFT-calculated local moments on Ni single motif (FM-like), double motif (AFM-like), as well as spin-SQS PM configurations as a function of the thickness N of the MgO barrier layer. We see that as the coupling between the building block motifs decreases (the number N of barrier layers increases) the local moment saturates. On the other hand, the magnitude of magnetic moments on Ni atoms in the FM NiO motif increases and saturates as the barrier thickness increases, i.e., for isolated FM-like building blocks. Interestingly, the bulk PM having no LRO has an average Ni magnetic moment magnitude similar to the FM-like and AFM-like motifs without LRO, which is between the values of bulk FM and bulk AFM. This suggests that the magnetic moment magnitude is mainly related to the LRO and the LRO with strong spin-up and spin-down (spin-up and spin-up) coupling reduces (enhances) the magnetic moment magnitudes as in bulk AFM (FM) structures.

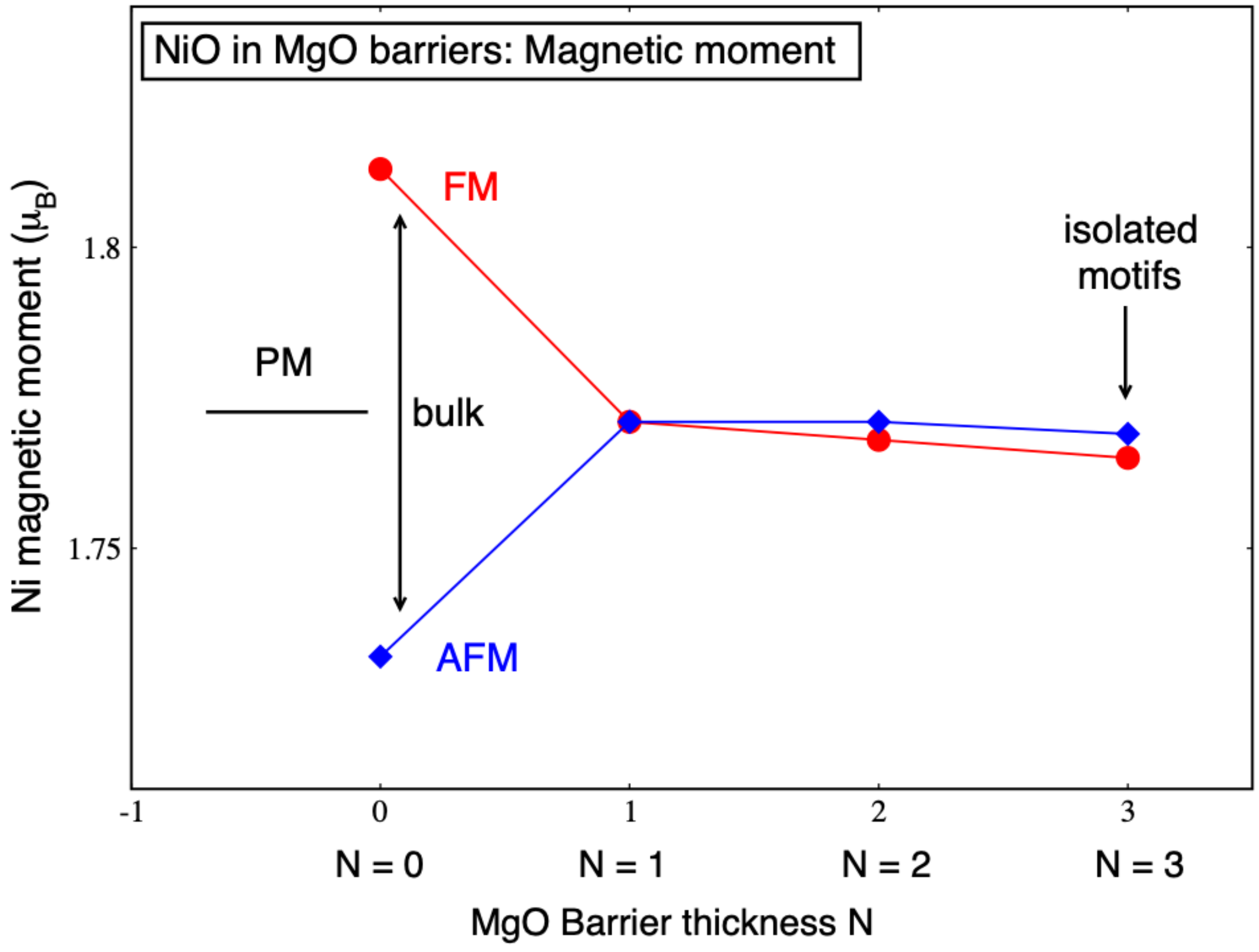

**Figure 4.** Local magnetic moments of FM and AFM of NiO as a function of the number of MgO barrier units N separating the individual motifs. N=0 corresponds to the bulk phases with fully interacting local motifs, whereas N=3 corresponds to isolated local motifs.

### D. The electronic structure of interacting motifs in bulk phases and in isolated motifs

Figure 5 compares the orbital projected density of states (PDOS) in bulk AFM NiO and isolated AFM-like spin motif embedded in MgO. Both configurations exhibit a finite band gap with closely similar orbital contributions, particularly near the band edges: The occupied (VBM) edge states contain both Ni-$t_{2g}$ and Ni-$e_g$ contributions, whereas the unoccupied (CBM) edge states are predominantly Ni ($e_g$) contributions. Upon isolation, the PDOS retains the characteristic band-edge orbital compositions of bulk AFM NiO. This preservation, together with the real-space Ni-d charge density distributions as shown in Fig. 2, supports the effectiveness of the MgO barrier in electronically isolating the spin motif while preserving its intrinsic electronic structure. The persistence of band gap in this limit (N=3 spin motif) indicates that LRO is not a prerequisites for gap formation and points to a local mechanism due to crystal-field splitting in combination of spin polarization. Symmetry-broken DFT captures this behavior without an explicit Hubbard *U* of dynamical correlation treatment.

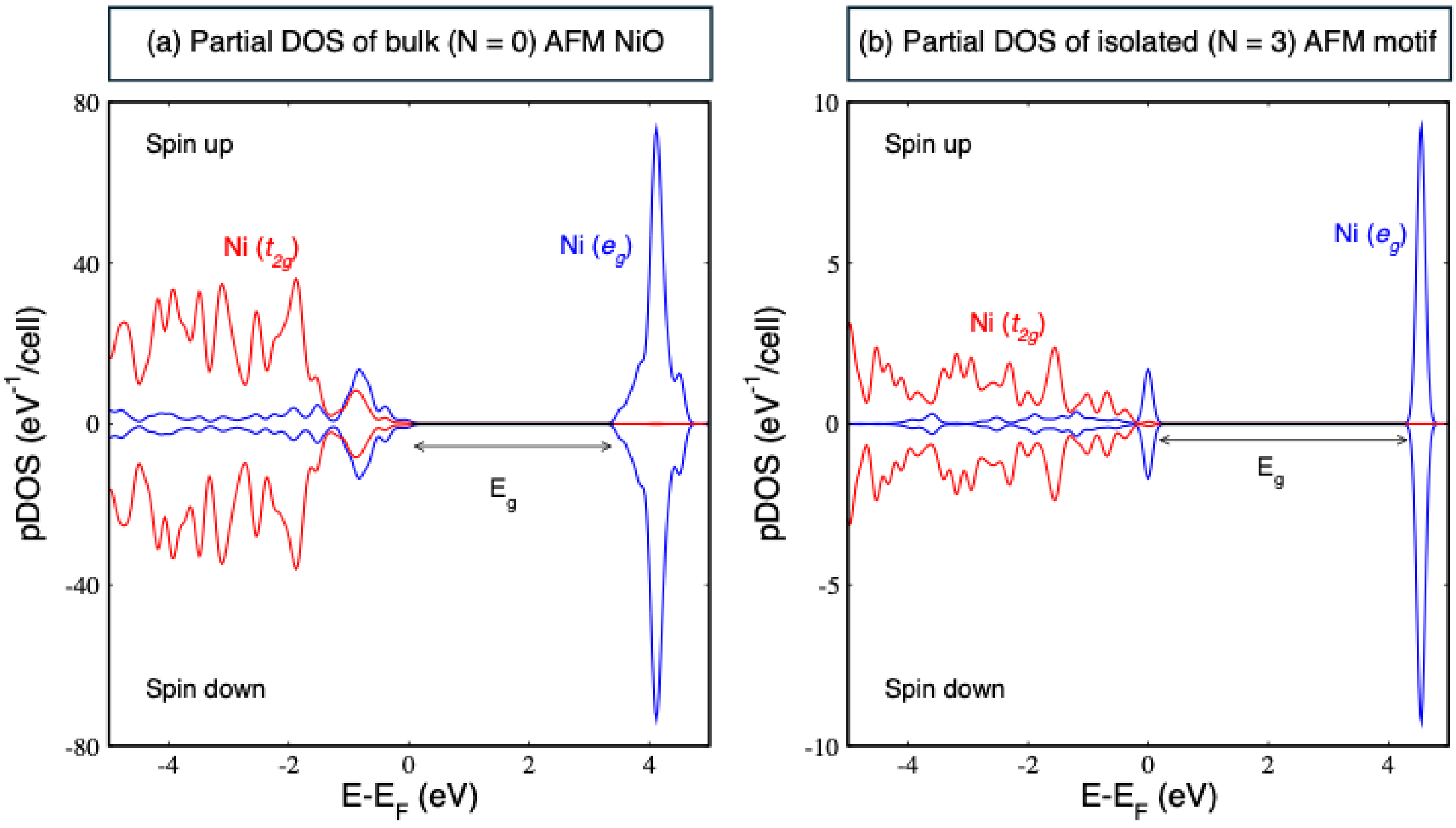

**Figure 5.** Projected density of states (PDOS) of (a) N=0 bulk NiO AFM and (b) N=3 AFM-like spin motif NiO-MgO, both having intrinsic band gaps. The O-p orbital contributions are removed for direct comparison of Ni-d orbitals in the two cases.

## IV. Discussions and summary

(i) *The resolution to the Mott-Slater controversy is provided:* The origin of the band gaps (metallic or insulating) in AFM, FM and PM phases obtained in first-principles DFT supercell calculations is neither LRO nor strong correlation. Instead, local *magneto-structural motifs* in the magnetic crystal—such as the magnetic atom with its surrounding nonmagnetic ligand atoms, may carry an intrinsic band gap that arises from symmetry breaking describable by the mean-field-like DFT interactions.

(ii) Trends in *band gaps of isolated building block motifs (large N) vs band gaps of bulk phases with interacting motifs (N=0):* The isolated local motifs (large N limit) of AFM-like and FM-like NiO have an *intrinsic* band gap, even larger than that of interacting motifs (bulk AFM phase, N=0), as shown in Fig. 3(a). This suggests that inter-motif interaction (specifically, LRO that will create level repulsion thus increase of gap) is not the reason for opening the band gap in isolated building blocks but rather reduces the band gap.

*(iii) Isolated local AFM motifs vs. bulk PM:* The calculated band edges of the NiO PM phase are compared to the AFM NiO and the NiO local motif in Fig. 3(b). The band gaps of AFM local motifs *are* considerably larger than the band gap of the bulk PM phase as the fluctuation of inter-motif coupling strengths in the PM phase reduces the band gap. Furthermore, the band gap of the bulk PM phase is much smaller than the AFM phase. These findings suggest that the interaction between the local motifs in both AFM and PM phases reduces the band gap.

*(iv) Isolated FM vs isolated AFM building-block motifs have gaps that converge to similar values:* The band alignment between isolated FM-like and isolated AFM-like configurations have the same CBM and VBM band edges, as shown in Fig. 3(b). This reflects the fact that the intrinsic band gap of local motifs exists due to the spin polarization in magnetic systems. The nearest-neighbor motif-coupling in opposite-spin $NiO_6$ octahedra in the AFM-like motif (only one spin-up motif couple with a spin-down motif in one direction) has negligible effects on the gap. Coincidently, as shown in Fig. 4, the nearest-neighbor motif-coupling in opposite-spin $NiO_6$ octahedra in the AFM motif has negligible effects on the magnetic moment magnitude.

*(v) The PM configuration has a slightly smaller gap than the AFM:* Transitioning from the well separated NiO configurations (N>0) to the non-separated NiO AFM structure (N=0), the band gap gradually *decreases,* mimicking the change of band gaps of quantum dots surrounded by barrier materials as the barrier materials shrinks and disappears. The slightly smaller band gap in bulk PM NiO compared with the bulk AFM NiO can be understood in terms of the distribution of local magnetic environments in the spin-disordered phase. These environments produce variations in the energies of the occupied and unoccupied states associated with individual spin motifs due to complex motif couplings. The highest occupied and lowest unoccupied states across these environments define the band edges, thereby narrowing the PM gap relative to its AFM counterpart.

## Acknowledgments

This work was supported by the U.S. Department of Energy, Office of Science, Basic Energy Sciences, Materials Sciences and Engineering Division under Grant No. DE-SC0010467, and in part by the RASEI Seed Program at University of Colorado Boulder. This work used resources from the National Energy Research Scientific Computing Center (NERSC), which is supported by the Office of Science of the U.S. Department of Energy.

## References

[1] N. F. Mott, The basis of the electron theory of metals, with special reference to the transition metals, P. Phys. Soc. Lond. A 62, 416 (1949).

[2] J. C. Slater, Magnetic effects and the Hartree-Fock equation, Phys. Rev. 82, 538 (1951).

[3] J. A. Alonso, M. J. Martínez-Lope, M. T. Casais, M. A. G. Aranda, and M. T. Fernández-Díaz, Metal-insulator transitions, structural and microstructural evolution of $RNiO_3$ (R = Sm, Eu, Gd, Dy, Ho, Y) perovskites: Evidence for room-temperature charge disproportionation in monoclinic $HoNiO_3$ and $YNiO_3$, J. Am. Chem. Soc. 121, 4754 (1999).

[4] J. Höwing, T. Gustafsson, and J. O. Thomas, Low-temperature structure of $V6O_{13}$, Acta Crystallogr B 59, 747 (2003).

[5] T. Toriyama, T. Nakayama, T. Konishi, and Y. Ohta, Charge and orbital orderings associated with metal-insulator transition in $V_6O_{13}$, Phys. Rev. B 90, 085131 (2014).

[6] S. Catalano, M. Gibert, J. Fowlie, J. Iñiguez, J. M. Triscone, and J. Kreisel, Rare-earth nickelates $RNiO_3$: thin films and heterostructures, Rep. Prog. Phys. 81, 046501 (2018).

[7] F. Serrano-Sánchez, J. L. Martinez, F. Fauth, and J. A. Alonso, On the lack of monoclinic distortion in the insulating regime of $EuNiO_3$ and $GdNiO_3$ perovskites by high-angular resolution synchrotron X-ray diffraction: a comparison with $YNiO_3$, Dalton T. 50, 7085 (2021).

[8] N. F. Mott and R. Peierls, Discussion of the paper by De Boer and Verwey, P. Phys. Soc. 49, A72 (1937).

[9] A. Carta, I. Timrov, P. Mlkvik, A. Hampel, and C. Ederer, Explicit demonstration of the equivalence between DFT+U and the Hartree-Fock limit of DFT+DMFT, Phys. Rev. Res. 7, 013289 (2025).

[10] D. Gambino, O. I. Malyi, Z. Wang, B. Alling, and A. Zunger, Density functional description of spin, lattice, and spin-lattice dynamics in antiferromagnetic and paramagnetic phases at finite temperatures, Phys. Rev. B 106, 134406 (2022).

[11] O. I. Malyi, X. G. Zhao, and A. Zunger, Insulating band gaps both below and above the Neel temperature in d-electron $LaTiO_3$, $LaVO_3$, $SrMnO_3$, and $LaMnO_3$ perovskites as a symmetry-breaking event, Phys. Rev. Mater. 7, 074406 (2023).

[12] A. Zunger, S. H. Wei, L. G. Ferreira, and J. E. Bernard, Special quasirandom structures, Phys. Rev. Lett. 65, 353 (1990).

[13] J.-X. Xiong, X. Zhang, and A. Zunger, Symmetry breaking forms split-off flat bands in quantum oxides controlling metal versus insulator phases, Phys. Rev. B 111, 035154 (2025).

[14] J.-X. Xiong, X. Zhang, and A. Zunger, Spin disorder competing with positional symmetry breaking governs the metal-insulator behavior in oxide paramagnets, arXiv:2606.14624 (2026).

[15] Z. Wang, O. I. Malyi, X. Zhao, and A. Zunger, Mass enhancement in 3d and s-p perovskites from symmetry breaking, Phys. Rev. B 103, 165110 (2021).

[16] J.-X. Xiong, X. Zhang, and A. Zunger, Role of magnetic and structural symmetry breaking in forming the Mott insulating gap in $Nb_3Cl_8$, Phys. Rev. B 111, 155122 (2025).

[17] G. Kotliar, S. Y. Savrasov, K. Haule, V. S. Oudovenko, O. Parcollet, and C. A. Marianetti, Electronic structure calculations with dynamical mean-field theory, Rev. Mod. Phys. 78, 865 (2006).

[18] V. M. Goldschmidt, T. Barth, D. Holmsen, G. Lunde, and W. Zachariasen, Geochemische Verteilungsgesetze der Elemente, Skrifter utgitt av det Norske Videnskaps-Akademi i Oslo 1: Matematisk-Naturvidenskapelig Klasse 1926, 2 (1926).

[19] E. Ressouche, N. Kernavanois, L. P. Regnault, and J. Y. Henry, Magnetic structures of the metal monoxides NiO and CoO re-investigated by spherical neutron polarimetry, Phys. B Condens. Matter 385, 394 (2006).

[20] G. Kresse and J. Furthmüller, Efficiency of ab-initio total energy calculations for metals and semiconductors using a plane-wave basis set, Comput. Mater. Sci. 6, 15 (1996).

[21] G. Kresse and J. Furthmüller, Efficient iterative schemes for ab initio total-energy calculations using a plane-wave basis set, Phys. Rev. B 54, 11169 (1996).

[22] G. Kresse and D. Joubert, From ultrasoft pseudopotentials to the projector augmented-wave method, Phys Rev B 59, 1758 (1999).

[23] B. Hammer, L. B. Hansen, and J. K. Nørskov, Improved adsorption energetics within density-functional theory using revised Perdew-Burke-Ernzerhof functionals, Phys. Rev. B 59, 7413 (1999).

[24] G. Trimarchi, Z. Wang, and A. Zunger, Polymorphous band structure model of gapping in the antiferromagnetic and paramagnetic phases of the Mott insulators MnO, FeO, CoO, and NiO, Phys. Rev. B 97, 035107 (2018).

[25] L. Zhang, P. Staar, A. Kozhevnikov, Y.-P. Wang, J. Trinastic, T. Schulthess, and H.-P. Cheng, DFT+DMFT calculations of the complex band and tunneling behavior for the transition metal monoxides MnO, FeO, CoO, and NiO, Phys. Rev. B 100, 035104 (2019).

[26] S. Lany, J. Osorio-Guillén, and A. Zunger, Origins of the doping asymmetry in oxides: Hole doping in NiO versus electron doping in ZnO, Phys. Rev. B 75, 241203 (2007).

[27] S. D. Singh, M. Nand, R. S. Ajimsha, A. Upadhyay, R. Kamparath, C. Mukherjee, P. Misra, A. K. Sinha, S. N. Jha, and T. Ganguli, Determination of band offsets at strained NiO and MgO heterojunction for MgO as an interlayer in heterojunction light emitting diode applications, Appl. Surf. Sci. 389, 835 (2016).